\documentstyle[12pt]{article}
\def\PRL #1 #2 #3{{\sl Phys.  Rev.  Lett.} {\bf#1} (#2) #3}
\def\NPB #1 #2 #3{{\sl Nucl.  Phys.} {\bf B #1} (#2) #3}
\def\NPBFS #1 #2 #3 #4{{\sl Nucl.  Phys.} {\bf B #2} [FS#1] (#3) #4}
\def\CMP #1 #2 #3{{\sl Commun.  Math.  Phys.} {\bf #1} (#2) #3}
\def\PRD #1 #2 #3{{\sl Phys.  Rev.} {\bf D #1} (#2) #3}
\def\PLA #1 #2 #3{{\sl Phys.  Lett.} {\bf #1A} (#2) #3}
\def\PLB #1 #2 #3{{\sl Phys.  Lett.} {\bf B #1} (#2) #3}
\def\JMP #1 #2 #3{{\sl J.  Math.  Phys.} {\bf #1} (#2) #3}
\def\PTP #1 #2 #3{{\sl Prog.  Theor.  Phys.} {\bf #1} (#2) #3}
\def\SPTP #1 #2 #3{{\sl Suppl.  Prog.  Theor.  Phys.} {\bf #1} (#2) #3}
\def\AoP #1 #2 #3{{\sl Ann.  of Phys.} {\bf #1} (#2) #3}
\def\PNAS #1 #2 #3{{\sl Proc.  Natl.  Acad.  Sci.  USA} {\bf #1} (#2)
#3}
\def\RMP #1 #2 #3{{\sl Rev.  Mod.  Phys.} {\bf #1} (#2) #3}
\def\PR #1 #2 #3{{\sl Phys.  Reports} {\bf #1} (#2) #3}
\def\AoM #1 #2 #3{{\sl Ann.  of Math.} {\bf #1} (#2) #3}
\def\UMN #1 #2 #3{{\sl Usp.  Mat.  Nauk} {\bf #1} (#2) #3}
\def\FAP #1 #2 #3{{\sl Funkt.  Anal.  Prilozheniya} {\bf #1} (#2) #3}
\def\FAaIA #1 #2 #3{{\sl Functional Analysis and Its Application}
{\bf #1} (#2) #3}\def\BAMS #1 #2 #3{{\sl Bull.  Am.  Math.  Soc.} {\bf
#1} (#2) #3}
\def\TAMS #1 #2 #3{{\sl Trans.  Am.  Math.  Soc.} {\bf #1} (#2) #3}
\def\InvM #1 #2 #3{{\sl Invent.  Math.} {\bf #1} (#2) #3}
\def\LMP #1 #2 #3{{\sl Letters in Math.  Phys.} {\bf #1} (#2) #3}
\def\IJMPA #1 #2 #3{{\sl Int.  J.  Mod.  Phys.} {\bf A #1} (#2) #3}
\def\AdM #1 #2 #3{{\sl Advances in Math.} {\bf #1} (#2) #3}
\def\RMaP #1 #2 #3{{\sl Reports on Math.  Phys.} {\bf #1} (#2) #3}
\def\IJM #1 #2 #3{{\sl Ill.  J.  Math.} {\bf #1} (#2) #3}
\def\APP #1 #2 #3{{\sl Acta Phys.  Polon.} {\bf #1} (#2) #3}
\def\TMP #1 #2 #3{{\sl Theor.  Mat.  Phys.} {\bf #1} (#2) #3}
\def\JPA #1 #2 #3{{\sl J.  Physics} {\bf A#1} (#2) #3}
\def\JSM #1 #2 #3{{\sl J.  Soviet Math.} {\bf #1} (#2) #3}
\def\MPLA #1 #2 #3{{\sl Mod.  Phys.  Lett.} {\bf A #1} (#2) #3}
\def\JETP #1 #2 #3{{\sl Sov.  Phys.  JETP} {\bf #1} (#2) #3}
\def\JETPL #1 #2 #3{{\sl Sov.  Phys.  JETP Lett.} {\bf #1} (#2) #3}
\def\PHSA #1 #2 #3{{\sl Physica} {\bf A #1} (#2) #3}
\def\CQG #1 #2 #3{{\sl Class.  Quantum Grav.} {\bf #1} (#2) #3}
\def\SJNP #1 #2 #3{{\sl Sov. J.  Nucl. Phys. (Yadern.Fiz.)} {\bf #1}
(#2) #3}
\def\a{\alpha}\def\b{\beta}\def\g{\gamma}\def\d{\delta}\def\e{\epsilon}
\def\E{\varepsilon}
\def\k{\kappa}\def\L{\Lambda}\def\s{\sigma}\def\S{\Sigma}
\def\Th{\Theta}\def\om{\omega}\def\Om{\Omega}\def\G{\Gamma}
\def\und{\underline}

\newcommand{\p}[1]{(\ref{#1})}
\newcommand{\plabel}{\label}
\begin{document}
\renewcommand{\thefootnote}{\fnsymbol{footnote}}
\setcounter{page}0
\thispagestyle{empty}
\begin{flushright}
{\bf 
hep-th/0010044} \\
{\bf FTUV-00-1006}\\ 
{\bf 2000, October 6}\\ 
{\bf 2000, November 15}
\end{flushright}

\vspace{1.5cm}

\begin{center}
{\LARGE
Lorentz harmonics and superfield action.
\\ $D=10$, $N=1$ superstring}

\vspace{1.5cm}
\renewcommand{\thefootnote}{\dagger}

{\bf Igor  Bandos$^{1,2}$, Tatyana Bandos$^{3,4}$}

\vspace{1.0cm}

{\it
$^{1}$Institute for Theoretical Physics,
NSC KIPT, \\ UA-61108 Kharkov, Ukraine, \\  \vspace{0.3cm}
$^{2}$ Departamento de Fisica Teorica \\
Universidad de Valencia, \\
E-46100 Burjassot (Valencia), Spain 
\\ \vspace{0.6cm}
$^{3}$
Abdus Salam ICTP, Trieste, Italy \\
\vspace{0.3cm}
$^{4}$
B.I. Verkin Institute for Low Temperature Physics, Kharkov, Ukraine}

\vspace{1.5cm}

{\bf Abstract}
 \end{center}

\medskip

{\small

We propose a new version of the superfield action for a closed
$D=10,~N=1$
superstring where the Lorentz harmonics are used as auxiliary
superfields.
The incorporation of Lorentz harmonics into the superfield
action makes possible to obtain superfield constraints of
the induced worldsheet supergravity as equations of motion.
Moreover, it becomes evident that a so-called 'Wess-Zumino
part' of the superfield action is basically a Lagrangian form of the
generalized action principle.
We propose to use the second Noether theorem
to handle the essential terms in the transformation lows of hidden
gauge symmetries, which remove dynamical degrees of freedom from the
Lagrange 
multiplier superfield.
}

\vspace{1.0cm} 

\setcounter{page}1
\renewcommand{\thefootnote}{\arabic{footnote}} \setcounter{footnote}0
\setcounter{page}0

\newpage

\bigskip


\section*{Introduction}

Recently a  new interest to the superfield description of superbranes
is witnessed \cite{Iv,pst2}.
The superfield actions for superbranes  might be useful in a search
for new superconformal theories
\cite{AdSbr}.

The superfield description of the Brink-Schwarz superparticle was
discovered
by Sorokin, Tkach, Volkov and Zheltukhin (STVZ) in Refs.
\cite{STV,VZ,STVZ}
(see \cite{Dima} for a nice review). In \cite{STV} and \cite{STVZ} the
geometrical origin of $\kappa$-symmetry as local worldline supersymmetry
has been established for $D=3,4$ and $D=10$ superparticles.
The STV--like actions have been also constructed  for
superstrings and some superbranes with not more than 16 target space
supersymmetries and 8 worldvolume supersymmetries
(\cite{Berk}--\cite{BMS}
and Refs. in \cite{Dima}).
So, the action for a $D=10$, $N=1$ closed superstring (heterotic
superstring
without heterotic fermions) was built in \cite{GHDS}
\footnote{The problem of  superfield description of the heterotic fermions 
is the separate subject considered in Refs. \cite{ST,H94,IS}.}.
The purpose of this paper is
to present another form of the action for such an object where Lorentz
harmonics \cite{Sok,B90,gds,BZ,bpstv}
(see also
\cite{niss,kallosh,Wiegmann}) are used.

The main advantage of this action is that the constraints of $d=2$,
$n=(8,~0)$ induced supergravity can be derived  as equations of motion,
while in the original approach \cite{STV,GS91,GHDS,Dima}
these constraints are imposed 'by hands'.
Such property might be useful for the consideration of STV-like actions
for higher superbranes. Moreover, the construction of such an action can be
regarded as a completion of the program of passing from the component
(Green--Schwarz)
action to the superfield action through an intermediate step of
the generalized
action principle \cite{bsv}
(see \cite{B98} for general consideration).

The superfield  action involves two Lagrange multiplier superfields.
To convince oneself that they do not carry dynamical degrees of freedom,
one 
should find some hidden gauge symmetries. Usually this is a nontrivial
task. However, here we propose a shortcut, namely, we shall demonstrate
how 
the second  Noether theorem provides the possibility of finding the
basic, essential terms in the transformation lows of the hidden gauge
symmetries by studying an interdependence of equations of motion.

We hope that the methods of the present paper will be useful in the
study of superfield formulations of higher superbranes in relatively
low dimensions D=4,5,6 as well as
for the investigation of  STV--like actions with $16$ supersymmetries 
in $D=11$ and $D=10$ type II 
superspaces \cite{PT,BS} with the aim to clarify the field content of
corresponding 'spinning superbrane' models 
\footnote{We call such  dynamical systems 'spinning superbranes' as 
i) they have both the worldsheet and space--time supersymmetry manifest 
and ii) they contain additional 
(in comparison with the Green--Schwarz superstring 
and $D=11$ supermembrane \cite{M2}) 
dynamical degrees of freedom \cite{bpstv}. 
Due to these properties  they  can be regarded as some extended counterparts 
of the 'spinning superparticle'  models \cite{SSp}. }.

The paper is organized as follows.
Section 1 is devoted to a brief review of the STVZ approach to
superparticles
and $N=1$ superstrings. In Section 2 we recall the generalized action
principle for $D=10$, $N=1$ superstring \cite{bsv},
derive the superfield equations
of motion and present some basic relations which are
useful for the consideration of
the  superfield action. The new version of the superfield action for the
$D=10$, $N=1$ superstring is presented in Section 3. The generation of
the worldvolume supergravity constraints by superfield equations of motion
which follow from the superfield action is the subject of Section 4. The
superfield equations are investigated in Section 5. Then,
in Section 6, we use the second Noether theorem to find the
gauge symmetries of the
superfield action, including hidden gauge symmetries which act on the
Lagrange multiplier superfields.
These symmetries are used in Section 7 to prove the pure auxiliary
nature
of the Lagrange multiplier superfields and to derive the component
equations of motion. Some technical details are collected in Appendix.

\section{STVZ approach to superparticle and $N=1$  superstring.
Brief review.}

 The STVZ approach to the description of the Brink-Schwarz superparticle
\cite{STV,STVZ,Dima} is based on the consideration of an embedding
of the world sheet superspace
\begin{equation}\plabel{2}
{\cal M}^{(1|n)}=(\tau, \eta^{q}) \equiv (\tau^{++}, \eta^{+q}), 
\qquad q=1, \ldots n
\end{equation}
into the D-dimensional target superspace ($D=3,4,6,10$, $N=1$)
\begin{equation}\plabel{3}
\underline{{\cal M}}^{(D|2(D-2))}=(X^{\underline{m}},
\Theta^{\underline{\mu}}),  \qquad
\end{equation}
$$
\underline{m}=0,~1,\ldots,(D-1) \qquad
\underline{\mu}=0,~1,\ldots,2(D-2).
$$
In \p{2} the sign indices $++$, $+$ of the worldline coordinates
$\tau^{++}, \eta^{+q}$
denote their scaling dimension (see below). Their usage is instructive
in
some places (see e.g. Eq.\p{D+q}).

The embedding
\begin{equation}\plabel{1}
{\cal M}^{(1|n)} \rightarrow  \underline{{\cal M}}^{(D|2(D-2))}
\end{equation}
can be defined locally by coordinate superfields
\begin{equation}\plabel{4}
X^{\underline{m}}=\hat{X}^{\underline{m}}(\tau^{++},~\eta^{+q})
=\hat{x}^{\underline{m}}(\tau)+\eta^{+q}~\chi_{+q}^{\underline{m}}(\tau)+
\ldots
+(\eta^{+q})^n~S^{\underline{m}}_{[+n]}(\tau ), \qquad
\end{equation}
$$
\Theta^{\underline{\mu}}=\hat{\Theta}^{\underline{\mu}}(\tau^{++},
\eta^{+q})=
\hat{\theta}^{\underline{\mu}}(\tau)+\eta^{+q}~{\lambda}_{+q}^
{\underline{\mu}I}(\tau) +
\ldots + (\eta^{+q})^{n}{\S}_{[+n]}^{\underline{\mu}I}(\tau ).
$$
Here and in what follows we use the compact notations for higher degrees of the
Grassmann variables
\begin{equation}\plabel{n}
(\eta )^n  \equiv
{1 \over n!} \e_{q_1 \ldots q_n }
\eta^{+q_1} \ldots
\eta^{+q_n}, \qquad
(\eta )^{n-1}_q  \equiv
{1 \over (n-1)!} \e_{qq_1 \ldots q_{n-1} }
\eta^{+q_1} \ldots
\eta^{+q_{n-1}}. \qquad
\end{equation}

The most complete description, where all $\k$--symmetries are replaced
by worldsheet
supersymmetries, is achieved when the number of worldvolume Grassmann
coordinates $\eta^{+q}$ is half the number of target space
Grassmann coordinates $\Th^{\underline{\mu}}$ (for  $D=3,4,6,10$,
$N=1$,
$\underline{\mu}=1,2, \ldots , 2(D-2)$, $q=1,2, \ldots , (D-2))$.

To describe the Brink--Schwarz superparticle (but not the so--called
spinning
superparticle model \cite{SSp} containing additional degrees of
freedom) the embedding (\ref{1}) should be
subject to the constraints
\begin{equation}\plabel{6}
\hat{\Pi}^{\underline{m}}_{+q}=D_{+q}\hat{X}^{\underline{m}}
(\tau, \eta) -
iD_{+q}\hat{\Theta}^{\underline{\mu}I} \G_{\underline{\mu}
\underline{\nu}}^{\underline{m}}~\hat{\Theta}^{\underline{\nu}I}
(\tau, \eta)
=0, \qquad
\end{equation}
where
\begin{equation}\plabel{D+q}
D_{+q}=\partial_{+q}+2i\eta^{+q}\partial_{++}, \qquad
\partial_{++}\equiv
{\partial \over \partial \tau^{++}}, \qquad
~\partial_{+q}\equiv
{\partial \over \partial \eta^{+q}}
\end{equation}
and
$\G_{\underline{\mu}\underline{\nu}}^{\underline{\mu}}=\G_{\underline{\nu}
\underline{\mu}}^{\underline{\mu}}$ 
are 10-dimensional $16\times16$
$\g$-matrices in Majorana-Weyl representation.
Eq.(\ref{6}) was called the geometrodynamic condition  \cite{GS91} or
the basic superembedding equation \cite{hs96,Dima}.
It implies the vanishing of the fermionic components
of the pull-back (on the worldline superspace)
\begin{equation}\plabel{8}
\hat{\Pi}^{\underline{m}}=d\hat{X}^{\underline{m}}-
i~d\hat{\Theta}^{\underline{\mu}I}\G_{\underline{\mu}
\underline{\nu}}^{\underline{m}}~
\hat{\Theta}^{\underline{\nu}I} = w^{++} \hat{\Pi}^{\underline{m}}_{++}
+ d\eta^{+q} \hat{\Pi}^{\underline{m}}_{+q}
\end{equation}
of the covariant bosonic form (vielbein) of the flat target
superspace
\begin{equation}\plabel{7}
{\Pi}^{\underline{m}}=d{X}^{\underline{m}}-
id{\Theta}^{I}\G^{\underline{m}}{\Theta}^{I}.
\end{equation}
In Eq.(\ref{8}) the 1--forms
\begin{equation}\plabel{9}
w^{++}=d\tau^{++}-2id\eta^{+q}\eta^{+q}, \qquad d\eta^{+q}
\end{equation}
provide a basis (supervielbein) for the differential forms on the
worldline
superspace. This basis is
invariant under superconformal symmetry \cite{gds,GS91}
\begin{equation}\plabel{10}
\d \tau^{++}=\L^{++}(\tau,\eta)=2a^{++}(\tau)+4i\eta^{+q} \k^{+q}(\tau)+
{1\over 2}\eta^{+q}\eta^{+p}~a_{[qp]}(\tau) + 
\ldots , \qquad
\end{equation}
$$
\d~\eta^{+q}=-{i\over 4}D_{+q}\L^{++}(\tau,\eta)=\k^{+q}+
\eta^{+p}~(\d_{qp}\partial_{++}a^{++}(\tau)+a_{qp}(\tau))+\ldots .
$$
The components  $a^{++}(\tau)$,
$\k^{+q}(\tau),~a_{pq}(\tau)=-a_{qp}(\tau),
\ldots$ of the superfield  $\L^{++}(\tau,\eta)$ can be regarded as the
parameters of gauge symmetries of the STV action \cite{STV,GS91}
\begin{equation}\plabel{11}
S_{0}= \int d\tau \hat{d}^{n}\eta~P_{\underline{m}q}~
\Pi^{\underline{m}}_{+q},   \qquad
\end{equation}
This action is basically the geometrodynamic condition \p{6}
incorporated with the Lagrange multiplier superfield
$P_{\und{m}}(\tau,\eta)$ whose  weight
under the induced Weyl rescaling with the parameter
${1\over2}\partial_{++}\L^{++}(\tau,\eta)$
(scaling dimension) is equal to $[+(n-1)]$. (The scaling dimension of
the
superspace measure $d\tau \hat{d}^{n}\eta $ is $[1-n]$).

 When the geometrodynamic condition Eq.\p{6} has no dynamical
equations among its consequences,
the Lagrange multiplier
$P_{\und{m}q}(\tau,\eta)$  does not contain
superfluous dynamical degrees of freedom\footnote{
This happens for $D=3,4,6,10$, $N=1$ and $D=3,4,6(IIA)$, $N=2$
superparticle, see \cite{GS92,BMS}}.  After making use of the equations
of
motion
\begin{equation}\plabel{12}
D_{+q} P_{\und{m}q}(\tau,\eta)=0,
\qquad P_{\und{m}q} (\tau,\eta)~ D_{+q}
\hat{\Theta}^{\underline{\mu}I}(\tau,\eta)~\G_{\underline{\mu}
\underline{\nu}}^{\underline{m}}
=0, \qquad
\end{equation}
and an infinitely reducible gauge symmetry \cite{GS91}
\footnote{This gauge symmetry acts on the Lagrange multiplier superfield
only
$\d P_{\und{m}q}= D_{+p} (\S^{qpr} \G_{\und{m}} D_{+r}\hat{\Th })$.
Its superfield parameter is symmetric and traceless
$\S^{pqr\und{\mu}}(\tau,\eta)= \S^{(pqr)\und{\mu}}$,
$\S^{ppr\und{\mu}}=0$.}
it can
be reduced to the form
\begin{equation}\plabel{13}
P_{\und{m}q}= {1 \over
(n-1)!}\e_{q q_1\ldots q_{n-1}} \eta^{+q_{1}}\ldots
\eta^{+q_{n-1}}~p_{\und{m}}
\equiv
(\eta )^{n-1}_q p_{\und{m}}
\end{equation} where $p_{\und{m}}$  is a
light--like constant vector \footnote{Indeed,  due to Eq.\p{12} $
\partial_{++}p_{\und{m}}(\tau)=0 \quad \Rightarrow \quad
p_{\und{m}}=const, \quad p_{\und{m}} p^{\und{m}} =0.$}
which can be identified with the momentum of a massless superparticle.
For the cases  when the geometrodynamic condition  contains equations of
motion among its consequences, as it is for a $D=10$ type $II$~($N=2$)
and
for a $D=6$ type $IIB$ superparticle \cite{BMS}, the Lagrange multiplier
superfield contains additional dynamical degrees of freedom. Thus the
action Eq. \p{11} describes a spinning superparticle model \cite{SSp}
in these cases
\footnote{ An alternative approach, which was suggested in \cite{IvK},
consists in solving Eq. \p{6} in terms of unconstrained superfields
(``prepotentials'').  Then an action for these prepotentials can be
constructed.  The recent superfield formulations of gauge-fixed brane
actions
\cite{Iv,pst2} follow mainly this line.}.

 The STV action  \p{11}  has  been generalized for some branes
with  not more than $16$ target space supersymmetries and thus not more
than $8$  worldsheet supersymmetries ($\k$-symmetries)
\cite{GHDS,Berk,BMS,HRS99,bcsv,Dima}.
(So the barrier of $16$ supersymmetries, known from the attempts
to construct superfield actions for supersymmetric field theories in
terms
of unconstrained superfields \cite{GIKOS},  has not been surmounted yet
\footnote{See \cite{N,CGNN} for recent progress in  an
off-shell description of $D=10, 11$ supergravity by relaxing the
torsion constraints.}).

 The superfield
 action for the branes has an additional term which was called the
Wess-Zumino term in \cite{GHDS}.
For $N=1$, $D=3,4,6,10$
superstring with worldsheet superspace
${\cal M}^{(2|(D-2))}=(\xi^{++},\xi^{--},\eta^{+q})=(\zeta^M)$
($q=1\ldots
(D-2)$, $\xi^{\pm\pm}=\tau \pm\s$) and coordinate superfields $$
X^{\und{m}}=\hat{X}^{\und{m}}(\xi^{++},\xi^{--},\eta^{+q}), \qquad
\Th^{\und{\mu}}=\hat{\Th}^{\und{\mu}}(\xi^{++},\xi^{--},\eta^{+q})
$$
the action of Ref.  \cite{GHDS} reads
\begin{equation}\plabel{14}
S_{1}= \int d^2\xi~\hat{d}^{D-2}\eta~P_{\und{m}q}
\Pi^{\und{m}}_{+q}+
\int d^2\xi~\hat{d}^{D-2}\eta~{\cal P}^{MN}
(\hat{\tilde{{\cal L}}}_2-d~Y_1)_{NM},
\qquad
\end{equation}
The second term contains the Lagrange multiplier superfield
${\cal P}^{MN}=-(-1)^{NM}{\cal P}^{NM}$, the auxiliary worldsheet
superfield
$Y_{M}(\zeta^{N})$ ($Y_1=d\zeta^{M}~Y_{M}(\zeta)$) and the 2--form
$\hat{\tilde{{\cal L}}}_2={1\over 2}d\zeta^{M} \wedge
d\zeta^{M}~\tilde{{\cal L}}_{MN}$;
($\tilde{{\cal L}}_{MN}= -(-)^{MN} \tilde{{\cal L}}_{NM}$), which we call
Lagrangian form for a reason which became transparent below. 
The latter is
constructed from $\hat{X}^{\und{m}}$,
$\hat{\Th}^{\und{\mu}}$ and some auxiliary
superfield.
In flat target superspace the Lagrangian form of Ref.
\cite{GHDS} is essentially 
\begin{equation}\plabel{SGHDS}
\hat{\tilde{{\cal L}}}=\hat{B}_2 + e^{++}\wedge e^{--}
\Pi^{\und{m}}_{++} \Pi_{\und{m}--} - dY,
\end{equation}
 where $B_2=-i\Pi^{\und{m}} \wedge d\Th\G^{\und{m}}\Th $ is the flat
 superspace value of the NS-NS 2-form and $e^{++}, e^{--}$ are bosonic
 supervielbein forms of the worldsheet supergravity, which are either
 subject to some set of torsion constraints \cite{bcsv} or constructed
 from some prepotentials \cite{GHDS}. Remember that
 $\Pi^{\und{m}}_{\pm\pm} = e_{\pm\pm}^{~M} \Pi_M^{\und{m}}$ and thus the
 action depends on the inverse supervielbein components
 $e^{M}_{\pm\pm}$ as well.

  It is
important that the Lagrangian form \p{SGHDS} is closed on the surface of
the geometrodynamic equation $\Pi_{+q}^{\und{n}}=0$
(this property was called Weyl triviality in
\cite{Tonin})
\begin{equation}\plabel{WT}
d\hat{\tilde{{\cal
L}}}_2\vert_{\Pi_{+q}^{\und{n}}=0}=0
\end{equation}
Due to this property
the equations of motion
\begin{equation}\plabel{dY}
\d S/\d {\cal
P}^{MN}=0 \qquad \Rightarrow \qquad \hat{\tilde{{\cal L}}}_{2}=dY_{1}
\end{equation}
can be regarded as {\sl nondynamical} equations for $Y_{1}$
\footnote{Indeed, as $\d Y_{1}=d
Y_{0}$ is a gauge symmetry of the model, Eq.\p{dY} completely determines 
 $Y_1$ in terms of closed form ${\tilde{\cal L}}_2$:
 after gauge fixing the solution does not involve {\sl any} indefinite
 constants. This means that the field $Y_1= d\xi^M Y_M$ has no
 independent degrees of freedom on the mass shell.}.
Then the equations of motion
$$
\d S/\d Y^{M}=0 \qquad \Rightarrow
\qquad (-)^{N}\partial_{N}{\cal~P}^{NM}=0 $$
and a set of gauge
symmetries provide us with the possibility of fixing a gauge
where the Lagrange multiplier ${\cal~P}^{NM}$ acquires the form
\begin{equation}\plabel{Pg}
{\cal P}^{NM}={1\over (D-2)!}
\e_{q_1\ldots~q_{(D-2)}}\eta^{+q_1}~ \ldots\eta^{+q_{(D-2)}}T
\equiv (\eta)^{D-2} T,
\qquad  \partial_{++}T=0.
\end{equation}
The constant $T$ has the meaning of a superstring tension.

The second Lagrange multiplier $P_{{\und{m}}q}$ does not
contain additional degrees of freedom. This can be motivated
starting from the observation that
the geometrodynamic condition
\begin{equation}\plabel{15}
\hat{\Pi}_{+q}^{\und{m}}= \nabla_{+q} ~\hat{X}^{\und{m}}-
i~\nabla_{+q} ~\hat{\Th}~\G^{\und{m}}\hat{\Th}=0   \qquad
\end{equation}
has no dynamical consequences. Thus it can be regarded as a pure
algebraic
equation. The Lagrange multiplier introduced to incorporate an
algebraic equation, as it is intuitively clear, should not carry
dynamical
degrees of freedom \cite{IS}.

In Eqs.\p{14}, \p{15} we should use the covariant derivatives of $d=2$
supergravity
$\nabla_{+q} =e_{+q}^{M} \partial_M$,
$\nabla_{\pm\pm} =e_{\pm\pm}^{M} \partial_M$,
but not the flat ones $D_{+q}=\partial_{+q}+2~i \eta^{+q}\partial_{++},
~\partial_{++},
~\partial_{--}$.
Thus  it is necessary  either to impose  the constraints
of $d=2$ supergravity or use their solution.
This is not a serious
problem for the
$N=1$, $D=10$ superstring, because $n=(8,0)$ $d=2$ supergravity is
conformally flat and does not contain dynamical degrees of freedom.
However, it might be a problem for the construction of the STV--like
actions for other branes with $p>1$ and $n\leq 8$ worldline
supersymmetries.

Here we present a reformulation of the model
\p{14} where the supergravity constraints occur as consequences 
of equations of motion. 
It differs from \p{14} by the choice of the 2-form ${\cal L}_2$.
Instead of  the components of the worldsheet
bosonic supervielbein forms $e_M^{\pm\pm}(\zeta )$
and $e^M_{\pm\pm}(\zeta )$ we use auxiliary light--like
10--vector superfields $U_{\und{m}}^{++}$, $U_{\und{m}}^{--}$
($U_{\und{m}}^{++} U^{\und{m}++}=0, U_{\und{m}}^{--} U^{\und{m}--}=0)$),
normalized by $U_{\und{m}}^{++} U^{\und{m}--}=2$, which can be regarded
as
Lorentz harmonics \cite{Sok}.  Moreover, our form $\hat{{\cal L}}_2$
is nothing but
{\sl the Lagrangian form of the generalized action} \cite{bsv}.

In the next section we recall main properties of the generalized
action and derive some formulae required for the development of the
superfield formalism.

\bigskip

\section{Generalized action and superfield description}
\centerline{\Large \bf of $N=1$, $D=10$ closed superstring}

\bigskip

 The generalized action \cite{bsv} for the $N=1$ $D=10$ superstring
\begin{equation}\plabel{2.1}
S= \int_{{\cal M}^{2}}  \hat{{\cal L}}_2   \qquad
\end{equation}
is an integral of a Lagrangian $2-$form ${\cal L}_2$ over an arbitrary
bosonic surface ${\cal M}^{2}=(\xi^{--}, \xi^{++}, \eta^{+q}(\xi))$
in the worldsheet superspace $\S^{(2|8)}=(\xi^{--},
\xi^{++},\eta^{+q})$.
The Lagrangian $2-$form
\begin{equation}\plabel{2.2}
{\cal L}_2={1\over 2}~E^{++}\wedge E^{--}-i~\Pi^{\und{m}}\wedge d\Th ~
\G_{\und{m}}\Th  \qquad
\end{equation}
is the sum of the traditional Wess-Zumino $2-$form
$B_2 =-i~\Pi^{\und{m}}\wedge d\Th ~\G_{\und{m}}\Th$
and a kinetic term constructed with the use of vector Lorentz harmonics
$U_{\und{m}}^{++},U_{\und{m}}^{--}$ \cite{Sok}
\begin{equation}\plabel{2.3}
E^{\pm\pm}=\Pi^{\und{m}}U^{\pm\pm}_{\und{m}}, \qquad
\Pi^{\und{m}}=dX^{\und{m}}-i~d\Th\G_{\und{m}}\Th ,   \qquad
\end{equation}

\subsection{$SO(1,9)/SO(1,1)\times SO(8)$ Lorentz harmonics}

The Lorentz harmonics $U^{\pm\pm}_{\und{m}}$ are elements of the Lorentz
group valued matrix
\begin{equation}\plabel{2.4}
U_{\und{m}}^{~\und{a}}=({1\over 2}(U^{++}_{\und{m}}+U^{--}_{\und{m}})),~
U_{\und{m}}^{I}, 
{1\over 2}(U^{++}_{\und{m}}-U^{--}_{\und{m}})) \in
SO(1,9)
\quad \Leftrightarrow \quad
U_{\und{m}}^{\und{a}}\eta^{\und{m}\und{n}}U_{\und{n}}^{\und{b}}=0
\end{equation}
It can be used to change the basis of cotangent superspace from
($\Pi^{\und{m}},d\Th^{\und{\mu}}$) to
\begin{equation}\plabel{2.5}
E^{\und{A}}=(E^{\und{a}}, E^{\und{\a}})
\end{equation}
where
\begin{equation}\plabel{2.6}
E^{\und{a}}= \left(
{1 \over 2} (E^{++} + E^{--}), ~E^{I}, ~{1 \over 2} (E^{++} -
E^{--})\right)=
\Pi^{\und{m}} U_{\und{m}}^{\und{a}}, \qquad
\end{equation}
$$E^{\pm\pm} =
\Pi^{\und{m}} U_{\und{m}}^{\pm\pm},  \qquad {} \qquad
E^{I} =
\Pi^{\und{m}} U_{\und{m}}^{I},$$
\begin{equation}\plabel{2.7}
E^{\und{\a}}=d\Th^{\und{\mu}}V_{\und{\mu}}^{~\und{\a}}
=(E^{+q}, E^{-}_{\dot{q}}),
\qquad ~~~~
E^{+q}=d\Th^{\und{\mu}}V_{\und{\mu}q}^{~+}, \qquad
E^{-}_{\dot{q}}=
d\Th^{\und{\mu}}V_{\und{\mu}\dot{q}}^{~-}.
\end{equation}
Eq.\p{2.7} contains the double--covering of the rotation matrix
$U_{\und{m}}^{\und{a}}$
\p{2.4}
\begin{equation}\plabel{2.8}
V_{\und{\mu}}^{\und{\a}}=(V^{+}_{\und{\mu}q},~V^{-}_{\und{\mu}\dot{q}}) 
~~
\in ~~Spin(1,9)
\end{equation}
It is defined by
\begin{equation}\plabel{2.9}
U_{\und{m}}^{\und{a}}\G_{\und{\mu}\und{\nu}}^{\und{m}}=
V_{\und{\mu}}^{\und{\a}}\G_{\und{\a}\und{\e}}^{\und{a}}
V_{\und{\nu}}^{\und{\e}},
\qquad {} \qquad 
U_{\und{m}}^{\und{a}}\G^{\und{\a}\und{\e}}_{\und{a}}=
V_{\und{\mu}}^{\und{\a}}\G^{\und{\mu}\und{\nu}}_{\und{m}}V_{\und{\nu}}^
{\und{\e}},
\end{equation}
and constructed from $8\times 16$ blocks 
$V^{+}_{\und{\mu}q},
V^{-}_{\und{\mu}q}$ which are called spinor
Lorentz harmonics
\cite{B90,gds,BZ}.

As the space tangent to the Lie group $SO(1,9)$ is isomorphic to
the Lie 
algebra $so(1,9)$ (which is the algebra of $10\times~10$ antisymmetric
matrices), the derivatives of $U_{\und{m}}^{a}$ are expressed in terms of 
antisymmetric {\sl Cartan forms}
\begin{equation}\plabel{2.10}
U_{\und{m}}^{\und{a}}dU^{\und{m}\und{b}}=\Om^{\und{a}\und{b}}=
- \Om^{\und{b}\und{a}}  \qquad \Leftrightarrow   \qquad
dU_{\und{m}}^{~\und{a}}=U_{\und{m}}^{~\und{b}}\Om^{~\und{a}}_{\und{b}}
\end{equation}
\begin{equation}\plabel{2.11}
V^{-1}{}^{~\und{\mu}}_{\und{\a}}dV^{~\und{\b}}_{\und{\mu}}=
{1\over
4}\Om^{\und{a}\und{b}}(\G_{\und{a}\und{b}})_{\und{\a}}^{~\und{\b}}
\qquad \Leftrightarrow  \qquad
d V_{\und{\mu}}^{~\und{\a}}={1\over 4}\Om^{\und{a}\und{b}}
V_{\und{\mu}}^{~\und{\b}}
(\G_{\und{a}\und{b}})_{\und{\a}}^{~\und{\b}}
\end{equation}
The splittings of $U$ and $V$ into the harmonic components
$U^{\pm\pm}_{\und{m}},  U^I_{\und{m}}$ and
$V^{+}_{\und{\mu}q},
V^{-}_{\und{\mu}q}$
are invariant under $SO(1,1)
\times SO(8)$ local transformations. The Cartan forms
$\Om^{\und{a}\und{b}}$
can be splitted into  blocks as well. These blocks 
are transformed as $SO(1,1)$ spin connections
\begin{equation}\plabel{2.12}
\omega={1\over 2}U_{\und{m}}^{--}dU^{\und{m}++},
\end{equation}
$SO(8)$ connections (gauge fields)
\begin{equation}\plabel{2.13}
A^{IJ}= U_{\und{m}}^{I}dU^{\und{m}J},
\end{equation}
and vielbein forms of the coset ${SO(1,9)\over SO(1,1)\times SO(8)}$
\begin{equation}\plabel{2.14}
f^{++I}=U^{++}_{\und{m}}dU^{\und{m}I},
\qquad {} \qquad  f^{--I}=U^{--}_{\und{m}}dU^{\und{m}I}.
\end{equation}

The Cartan forms \p{2.12} -\p{2.14} can be used to decompose
Eq.\p{2.10} as follows
\begin{equation}\plabel{2.15}
{\cal{D}}U^{++}_{\und{m}}=dU^{++}_{\und{m}}-U^{++}_{\und{m}}\omega=
f^{++I}U_{\und{m}}^{I},
\end{equation}
\begin{equation}\plabel{2.16}
{\cal{D}}U^{I}_{\und{m}}=dU^{I}_{\und{m}}+U^{I}_{\und{m}}A^{II}=
{1\over 2}U_{\und{m}}^{++}f^{--I}+{1\over 2}U_{\und{m}}^{--}f^{++I}.
\end{equation}
To decompose Eqs. \p{2.9},
an $SO(1,1)\times
SO(8)$ covariant representation of the Mayorana-Weyl $16\times16$
$\g$-matrices should be used \cite{BZ}.
In an appropriate representation
one obtains
\begin{equation}\plabel{2.17}
U^{++}_{\und{m}}\G^{\und{m}}_{\und{\mu}\und{\nu}}=
2~V^{+}_{\und{\mu}q}V^{+}_{\und{\nu}q},   \qquad
U^{--}_{\und{m}}\G^{\und{m}}_{\und{\mu}\und{\nu}}=
2~V^{-}_{\und{\mu}q}V^{-}_{\und{\nu}q},   \qquad
U^{I}_{\und{m}}\G^{\und{m}}_{\und{\mu}\und{\nu}}=
V^{+}_{\und{\mu}q}\g^{I}_{q\dot{q}}V^{-}_{\und{\nu}\dot{q}}+
V^{+}_{\und{\nu}q}\g^{I}_{q\dot{q}}V^{-}_{\und{\mu}\dot{q}}, \qquad
\end{equation}
$$
\delta_{qp}U^{++}_{\und{m}}=V^{+}_{q}\tilde{\G}_{\und{m}}V^{+}_{p},
\qquad
\d_{\dot{q}\dot{p}}U^{--}_{\und{m}}=V^{-}_{\dot{q}}\tilde{\G}_{\und{m}}
V^{-}_{\dot{p}},  \qquad
U^{i}_{\und{m}}\g^{i}_{q\dot{q}}=
V^{+}_{q}\tilde{\G}_{\und{m}}V^{-}_{\dot{q}}.
$$

\bigskip

Note that Eq. \p{2.4} implies 
$$U_{\und{m}}^{\und{a}}\eta^{\und{m}\und{n}}U_{\und{n}}^{\und{b}}=0
\quad \Leftrightarrow \quad 
$$
\begin{equation}\plabel{2.4'}
U^{++}_{\underline{a}} U^{++\underline{a}}=0, \qquad
U^{--}_{\underline{a}} U^{--\underline{a}}=0, \qquad
\end{equation}
\begin{equation}\plabel{2.4''}
U^{++}_{\underline{a}} U^{--\underline{a}}=2, 
\end{equation}
\begin{equation}\plabel{2.4'''}
U^{\pm\pm}_{\underline{a}} U^{i\underline{a}}=0, \qquad  
U^{i}_{\underline{a}} U^{j\underline{a}}=-\d^{ij}. \qquad 
\end{equation} 
Thus $V^{+}_{\und{\mu}q}$ and $V^{-}_{\und{\mu}\dot{q}}$ can be treated 
as 'square roots' of the light--like vectors 
$U^{++}_{\underline{a}}$ and 
$U^{--}_{\underline{a}}$ \p{2.4'} normalized by  \p{2.4''}. 

The integrability conditions for Eqs. \p{2.12}--\p{2.14}
provide us with the {\sl Maurer--Cartan equations} for the
Cartan forms
\begin{equation}\label{PC}
{\cal D} f^{\pm\pm I}
\equiv d f^{\pm\pm I} \mp f^{\pm\pm I} \wedge \omega +
f^{\pm\pm J} \wedge A^{JI} = 0,
\end{equation}
\begin{equation}\label{G}
d \omega = {1\over 2} f^{--I} \wedge f^{++I},
\end{equation}
\begin{equation}\label{R}
 {\cal F}^{IJ}  \equiv d A^{IJ} + A^{IK} \wedge A^{KJ}
= - f^{--[I} \wedge f^{++J]}.
\end{equation}

\bigskip

\subsection{External derivative of the Lagrangian $2$-form and
superfield equations of motion}

With the above notation it is straightforward to calculate a formal
external derivative of the Lagrangian $2$-form \p{2.2}
\begin{equation}\plabel{3.1}
d{\cal L}_2={1\over 2}~E^{I}\wedge (E^{--}\wedge f^{++I}-E^{++}\wedge
f^{--I}-
4iE^{+q}\wedge E^{-\dot{q}}\g^{I}_{q\dot{q}})
-2i~E^{-}_{\dot{q}}\wedge E^{-}_{\dot{q}}\wedge E^{++}.  \qquad
\end{equation}
The variation of the action \p{2.1}, \p{2.2} can be easily derived  from 
\p{3.1}
by making use of the seminal formula
\footnote{
Our notation for the external derivative and contraction of a $q$-form
\\
$
\Omega_{q}={1\over q!}dZ^{\und{M}_q}\wedge \ldots \wedge dZ^{\und{M}_1}
\Omega_{\und{M}_1 \ldots \und{M}_q}(Z)
$
is as follows
$$
d\Omega_{q}={1\over q!}dZ^{\und{M}_q}\wedge \ldots \wedge dZ^{\und{M}_1}
\wedge dZ^{\und{N}}\partial_{\und{N}}\Omega_{\und{M}_1 \ldots \und{M}_q}
(Z), \qquad
i_{\d}\Omega_{q}={1\over (q-1)!}dZ^{\und{M}_q}\wedge \ldots \wedge
dZ^{\und{M}_2}
\delta Z^{\und{M}_1}
\Om_{\und{M}_1,\und{M}_2 \ldots \und{M}_q} 
(Z) $$
}

\begin{equation}\plabel{3.2}
\delta {\cal L}_2=i_{\d}d{\cal L}_2+di_{\d}{\cal L}_2
\end{equation}
supplemented by the rules
\begin{equation}\plabel{3.3}
i_{\d}dZ^{\und{M}}=\delta \hat{Z}^{\und{M}}.
\end{equation}
The contractions of Cartan forms $i_{\d}f^{++I}$, $i_{\d}f^{--I}$ should
be
considered as parameters of independent variations of the Lorentz
harmonic
superfields $U$ and $V$.

The equations of motion which follow from generalized action are
\begin{equation}\plabel{3.4}
\hat{E}^I=\hat{\Pi}^{\und{m}}U_{\und{m}}^I=0, 
\end{equation}
\begin{equation}\plabel{3.5}
\hat{\Psi}_2{}^{+}_{\dot{q}} \equiv E^{++}\wedge E^{-}_{\dot{q}} = 0, 
\end{equation}
\begin{equation}\plabel{3.6}
\hat{M}_2^I=
{1 \over 2} E^{++}\wedge f^{--I}
- {1 \over 2} E^{--}\wedge f^{++I}-
2iE^{+}_{q}\wedge E^{-}_{\dot{q}}\g^{I}_{q\dot{q}}=0,   \qquad
\end{equation}

The key point is that Eqs.{\p{3.4}-\p{3.6}} can be regarded as
superfield
equations which {\sl are valid in the whole worldvolume superspace}
\cite{bsv}.  In such a treatment Eq. \p{3.4} (the basic superembedding
equation) is an equivalent form of geometrodynamic condition  \p{15}
\cite{bpstv}. It does not contain the fermionic equations of motion
\p{3.5} among its consequences.  Studying the integrability conditions
for Eq.\p{3.4}
\begin{equation}\plabel{3.5''}
{\cal{D}}\hat{E}^I=d\hat{E}^I+A^{IJ} \wedge \hat{E}^J=
{1\over 2} \hat{E}^{--}\wedge f^{++I} +
{1\over 2} \hat{E}^{++}\wedge f^{--I}
-2i\hat{E}^+_{q}\wedge \hat{E}^{-}_{\dot{q}}\g^{I}_{q\dot{q}}=0,
\qquad
\end{equation}
one finds that (see Section 6)
\begin{equation}\plabel{3.6'}
\hat{E}^{-}_{\dot{q}}= \hat{E}^{++}\psi^{~~~-}_{++\dot{q}}+
\hat{E}^{--}\psi^{~~~-}_{--\dot{q}},
\qquad
\end{equation}
while, as it is easy to see, Eq.\p{3.5} also implies
 $\psi^{-}_{--\dot{q}}=0$. The latter is just the fermionic
equation of motion.  Indeed, in the linear approximation 
it reduces to $\psi^-_{--\dot{q}}\approx \partial_{--}\Th^-_{\dot{q}}=0$,
where $\Th^-_{\dot{q}}=\Th^{\und{\mu}}V^-_{\und{\mu}\dot{q}}$.

On the other hand, one finds that the superembedding equation \p{3.4}
and fermionic equations of motion \p{3.5} determine the bosonic
equations
\p{3.5} completely. Indeed, on the surface of the superembedding
equation
\p{3.4}, where Eq. \p{3.5''} holds,
 the bosonic equation of motion \p{3.5} can be written in
the form
\begin{equation}\plabel{3.6''}
\hat{M}_2^I\vert_{\hat{E}^I=0} =
E^{++}\wedge f^{--I} -
4iE^{+}_{q}\wedge E^{-}_{\dot{q}}\g^{I}_{q\dot{q}}=0.           \qquad
\end{equation}
Then one easily finds that Eq. \p{3.6''} can be obtained  as a
derivative
of the fermionic equation \p{3.5}
\begin{equation}\plabel{3.61}
{\cal D} (\Psi_2)^{-}_{\dot{q}} ~\vert_{\hat{E}^I=0}=
-{1\over 2}
\hat{M}_2^I \wedge E^+_q \g^I_{q\dot{q}}.
\end{equation}

An important property of the Lagrangian $2-$form Eq.\p{2.2}  is that it
is closed on the surface of the superembedding equation Eq.\p{3.4}
\begin{equation}\plabel{3.7}
d\hat{{\cal L}}_2\vert _{\hat{E}^I=0} \equiv 0
\end{equation}
Indeed, the pull--back of the first term in Eq. \p{3.1} vanishes due to
the superembedding equation \p{3.4} while the second
one becomes zero when the consequence 
\p{3.6'} of Eq.\p{3.4} is taken into account. As Eq.\p{3.4} is an
equivalent form of the geometrodynamic condition \p{15}, one can say that
the Lagrangian form \p{2.2} of the generalized action \p{2.1} possesses
the property called Weyl triviality \cite{Tonin}.

Now we are ready to proceed with the superfield action.

\newpage 

\section{ New version of superfield action for $D=10$, $N=1$
closed superstring}

We propose the following superfield action functional for the $D=10$,
$N=1$ closed superstring
\begin{equation}\plabel{4.1}
S= \int d^2\xi \hat{d^8}\eta
\left(P^M_I \hat{E}^I_M+P^{MN}(\hat{{\cal L}}_2-dY_1)_{NM}\right).
\qquad \end{equation}
Here the superfield $\hat{E}^I_M$ emerges in the
decomposition of the pull-back of the bosonic vielbein form
$\hat{E}^I=\hat{\Pi}^{\und{m}}U^I_{\und{m}}$  on the worldsheet
superspace
$\sum^{(2|8)}=(\xi^m, \eta^q)=(\zeta^M)$
\begin{equation}\plabel{4.2}
\hat{E}^I=
d\zeta^M \hat{E}_M^I=\hat{\Pi}^{\und{m}}U^I_{\und{m}},
\qquad \Pi^{\und{m}}=dX^{\und{m}} - i d\Th \G^{\und{m}}\Th ,   \qquad
\end{equation}
$\hat{{\cal L}}_2$ is the Lagrangian form of the generalized action
\p{2.2}, but pulled-back onto the whole worldsheet superspace instead of
a bosonic surface in this superspace
\begin{equation}\plabel{4.3}
\hat{{\cal L}}_2={1\over 2}\hat{E}^{++}\wedge
\hat{E}^{--}-i\hat{\Pi}^{\und{m}}\wedge d\hat{\Th} \G_{\und{m}}\hat{\Th}
={1\over 2}d\zeta^M\wedge d\zeta^N \hat{{\cal L}}_{NM}. 
\qquad
\end{equation}
Thus
\begin{equation}\plabel{4.4}
(\hat{{\cal L}}_2)_{NM} = \hat{E}^{--}_{[N} \hat{E}^{++}_{M)}-2~i
\partial_{[N|}\hat{\Th}\G_{\und{m}}\hat{\Th}
\hat{\Pi}^{\und{m}}_{|\und{M})}
\end{equation}
where mixed brackets $[\ldots )$ denote graded
antisymmetrization with common weight unity,
e.g.  \begin{equation}\plabel{4.5} E^{--}_{[N} E^{++}_{M)}\equiv{1\over
2}(E^{--}_{N} E^{++}_{M}- (-)^{MN} E^{--}_ME^{++}_{N}).
\end{equation}
$P^{MN}=P^{[MN)}(\zeta)$ and
$P^{M}_I(\zeta)$ are Lagrange multiplier superfields and
\begin{equation}\plabel{4.6}
Y_1\equiv d\zeta^M Y_M (\zeta)
\end{equation}
is an auxiliary $1-$form superfield (cf. \cite{GHDS}).
(Thus $(dY_1)=d\zeta^M\wedge d\zeta^N \partial_N Y_M,~~$
$(dY_1)_{NM}=2\partial_{[N} Y_{M]}=
\partial_{N} Y_{M} - (-)^{NM}
\partial_{M} Y_{N}$).

\bigskip

\subsection{Variation of the action}

Usually the variation of the actions with superspace tensors is quite
involved due to many sign factors which appear in calculations. However, 
for the action Eq.\p{4.1} there exists a shortcut which is provided by
the
fact that all the expressions involved (except for the Lagrange multipliers) 
are expressed in terms of differential forms. Hence, to vary different
terms of the action one can use Eq. \p{3.2} and, then, extract the basic 
differential forms $d\zeta^M$,  $d\zeta^M \wedge d\zeta^N$.
For instance,
using the expression \p{3.5''} for the external derivative of the form
$\hat{E}^{I}=\Pi^{\und{m}}U_{\und{m}}^{I}$ and Eq. \p{3.2} one easily 
arrives at
\begin{equation}\plabel{4.9}
\d E^I_M=\partial_M (i_{\d}\Pi^{\und{m}})U_{\und{m}}^I- E^J_M i_\d
A^{JI} +
{1\over 2}E^{--}_M i_{\d}f^{++I}+{1\over 2}E^{++}_M i_{\d}f^{--I} -
\qquad
\end{equation}
$$
-2iE^{~+}_{Mq} \g^I_{q{\dot{q}}} i_{\d}E^{-}_{\dot{q}}
-2iE^{~-}_{M\dot{q}}\g^I_{q{\dot{q}}} i_{\d}E^{+}_{q}.
$$
Here and below we will skip the hat symbol $~\hat{\ldots}~$ from the
pull--backs of differential forms for the sake of shortness
(i.e. $\hat{E}^a$ reads now as $E^a$).

The variation of $({\cal{L}}_2)_{NM}$ can be found
in the similar way from Eqs.\p{3.1} and \p{3.2}. Making use of this
technique we arrive at the following expression for the variation of
superfield action \p{4.1}
\begin{equation}\plabel{4.10}
\d S= \int
d^2\xi \hat{d^8}\eta \Big[(\d P^M_I+P_J^M i_\d A^{JI})E^I_M+ \d
P^{MN}({\cal L}_2-dY_1)_{NM}-
\end{equation}
$$
- (-)^N\partial_N P^{NM}(\d Y_M-(i_{\d}{\cal L}_2)_M)+
$$
$$ + ({1\over 2}P^M_I E^{++}_M-P^{MN}E^{++}_N
E_M^I)i_{\d}f^{--I}+ ({1\over 2}P^M_I E^{--}_M+P^{MN}E^{--}_N
E_M^I)i_{\d}f^{++I}+
$$
$$
+ \left(-(-)^M(\partial_M P_I^M-A_M^{IJ}P^M_J)+
P^{MN}(f^{++I}_NE^{--}_M-f_N^{--I}E^{++}_M+4iE^{~-}_{N\dot{q}}
E^{~+}_{M{q}}(-)^M\g^I_{q\dot{q}}
\right) i_{\d}E^I-
$$
$$
- \left({1\over 2}
P^M_I f^{++I}_M +P^{MN}f^{++I}_N
E^I_M\right)i_{\d}E^{--}-
$$
$$
- {1\over 2}\left(P^M_I f^{--I}_M +8iP^{MN}\left(-(-)^M
E^{~-}_{N\dot{q}}E^{~-}_{M\dot{q}}+ {i\over 4}f^{--I}_N
E^I_M\right)\right)i_{\d}E^{++} -
$$
$$
-2i\left(P^M_I
E^{~+}_{Mq}\g^I_{q\dot{q}}-4P^{MN}
\left(E^-_{N\dot{q}}E^{++}_M-{1\over
2}E^{~+}_{Nq}\g^I_{q\dot{q}}E^I_M\right)\right) i_{\d}E^{~-}_{\dot{q}} -
$$
$$
-2i\left(P^M_I E^{~-}_{M\dot{q}}\g^I_{q\dot{q}}-2P^{MN}
E^{~-}_{N\dot{q}}\g^I_{q\dot{q}}(-)^ME^I_M\right)
i_{\d}E^{~+}_{q}\Big].   \qquad
$$

Let us turn to the equations of motion. The variations of the Lagrange
multipliers $P^M_I$, $P^{MN}$ produce the superembedding
equation \p{3.4} 
\begin{equation}\plabel{4.11}
\d P^M_I: \qquad
E^I_M=0 \quad \Rightarrow \quad   E^I
\equiv {\Pi}^{\und{m}}U^I_{\und{m}} =0
\qquad
\end{equation}
and the equation
\begin{equation}\plabel{4.12}
\d P^{MN}:  \qquad ({\cal L}_2-dY_1)_{NM}=0 \quad
\Rightarrow  \quad {\cal L}_2=dY_1.
\qquad \end{equation}
 Since the integrability conditions
for Eq.\p{4.12}
\begin{equation}\plabel{4.13}
d{\cal L}_2=0    \qquad
\end{equation}
are satisfied identically on the surface of the superembedding equation
\p{4.11} (see Eq.\p{3.7}), Eq.\p{4.12}
expresses the auxiliary  super--1--form $Y_1$ \p{4.6} through 
${\cal L}_2$ and thus is not a dynamical equation.
To arrive at  such a conclusion one should recall that
$\d Y_M=\partial_M f$ is a gauge symmetry of the action
(see Eq.\p{4.10}).

On the other hand, the above mentioned dependence of the integrability
conditions
\p{4.13} for Eq.\p{4.12} is nothing but the Noether identity which
reflects the presence of additional gauge symmetry with the basic
relation
$\d P^{MN}\equiv (-)^K\partial_K\Sigma^{KMN}$,
$\Sigma^{KMN}=\Sigma^{[KMN)}$.
The Noether identity for the $SO(8)$ symmetry is manifested by the fact that
the equations of motion which emerge as a result of the
variation $i_{\d}A^{IJ}$:
\begin{equation}\plabel{4.14}
i_{\d} A^{IJ}:  \qquad
P^M_I E^J_M -P^M_J E^I_M =0   \qquad
\end{equation}
are satisfied identically when Eq.\p{4.11} is taken into account.
We turn to further study of Noether identities and gauge symmetries
in Section 6.

The variation with respect to the auxiliary $2$-form superfield $\d
Y_M(\zeta)$ provides the equation
\begin{equation}\plabel{4.22}
(-)^N\partial_N P^{NM}=0.   \qquad
\end{equation}
The remaining equations
of motion on the surface of the superembedding condition \p{4.11}
have the
form

\begin{equation}\plabel{4.15}
i_{\d} f^{--I}: \qquad   {} \qquad  
P^M_I E^{++}_M =0,
\qquad \end{equation}

\begin{equation}\plabel{4.16}
i_{\d} f^{++I}:  \qquad   {} \qquad  
P^M_I
E^{--}_M =0,    \qquad \end{equation}
\begin{equation}\plabel{4.17}
i_{\d} E^{I}: \qquad   {} \qquad   (-)^M(\partial_M P^M_I- A^{IJ}_MP^M_J)=
\end{equation}
$$
  {} \qquad     {} \qquad   = P^{MN}
(f^{++I}_N E^{--}_M-f^{--I}_N E^{++}_M+
4i(-)^ME^{~-}_{N\dot{q}}E^{~+}_{Mq}\g^I_{q\dot{q}}),
$$

\begin{equation}\plabel{4.18}
i_{\d} E^{--}: \qquad   {} \qquad  P^M_I
f^{++I}_M =0,    \qquad
\end{equation}

\begin{equation}\plabel{4.19}
i_{\d} E^{++}:  \qquad   {} \qquad  P^M_I f^{--I}_M =
+4i P^{MN} (-)^M E_N{}^{-}_{\dot{q}} E_M{}^{-}_{\dot{q}},   \qquad
\end{equation}
\begin{equation}\plabel{4.20}
i_{\d} E^{-}_{\dot{q}}:
\qquad   {} \qquad  P^M_I\g^I_{q\dot{q}}E^{~+}_{Mq}= 4P^{MN}(-)^M
E^{~-}_{N\dot{q}}E^{++}_M,  \qquad \end{equation}
\begin{equation}\plabel{4.21}
i_{\d} E^{+}_q:  \qquad   {} \qquad  
P^M_IE^{~-}_{M\dot{q}}\g^I_{q\dot{q}}=0.  \qquad
\end{equation}

Eqs.\p{4.22}, \p{4.17} look like dynamical ones. However, as we will see 
in Section 6, there exist gauge symmetries which make possible to
gauge away the general solution of these equations. That means that the
Lagrange multiplier superfields do not contain dynamical degrees  of
freedom. Due to this fact the model described by the
action Eq.\p{4.1} is equivalent to the $D=10$, $N=1$ Green - Schwarz
superstring at the classical level.

\bigskip

\section{Generation of supergravity constraints}

The action \p{4.1}, \p{4.2}, \p{4.3} does not contain an intrinsic
worldsheet supervielbein
\begin{equation}\plabel{4.7.1} e^A=(e^{++},
e^{--}, e^{+q})=d\zeta^M e^A_M \qquad
\end{equation}
and intrinsic spin connections as independent
variables.  In this sense Eq.\p{4.1} can be regarded as a superfield
counterpart of Nambu-Goto action, while the original superfield action
\cite{GHDS} can be treated as a counterpart of the Brink--Di
Vecchia--Howe--Polyakov action.

However, to deal with the action, in particular to investigate equations
of
motion and to study  symmetries, one needs to assume that among the
pull--backs of the target space supervielbein there exists a set of two
bosonic and 8 fermionic forms which are linearly independent.
We assume that these forms are

\begin{equation}\plabel{4.7.2}
\hat{E}^{{\pm}{\pm}}=\hat{\Pi}^{\und{m}}U_{\und{m}}^{{\pm}{\pm}}=
d\zeta^M\hat{E}_M^{{\pm}{\pm}}(\zeta), \qquad
\hat{E}^{+q}_M=d\zeta^M \hat{E}_M^{+q}(\zeta) \qquad
\end{equation}
Thus the inverse blocks
\begin{equation}\plabel{4.7.3}
\hat{E}_{{\pm}{\pm}}^{~~M},\hat{E}_{+q}^{~M}, \qquad
\end{equation}
do exist and can be defined by
\begin{equation}\plabel{4.7.4}
\hat{E}_{+q}^M\hat{E}^{+p}_M=\d^p_q, \qquad
\hat{E}_{+q}^M~\hat{E}^{{\pm}{\pm}}_M~=0,  \qquad
\end{equation}
$$
\hat{E}_{{\pm}{\pm}}^M\hat{E}^{+q}_M~=0, \qquad
\hat{E}_{+q}^M~\hat{E}^{{\pm}{\pm}}_M=0,  \qquad
\hat{E}_{++}^M~\hat{E}^{--}_M=0=\hat{E}_{--}^M~\hat{E}^{++}_M. \qquad
$$

Actually, by this step we implicitly  assumed
that the worldsheet supervielbein can be induced by embedding in
accordance with the relations
\begin{equation}\plabel{4.2.8}
e^{\pm\pm} =
\hat{E}^{{\pm}{\pm}}\equiv
\hat{\Pi}^{\und{m}}U_{\und{m}}^{{\pm}{\pm}},
\qquad
e^{+q} =
\hat{E}^{+q}_M \equiv
d\hat{\Th}^{\und{\mu}}V_{\und{\mu}q}^{~+}.
\end{equation}
To complete the description of the worldsheet geometry one can choose
the worldsheet spin connection and the $SO(8)$ gauge field to be equal
to the Cartan forms \p{2.12}, \p{2.13}

\begin{equation}\plabel{cCf}
 \om = {1 \over 2}
U_{\und{m}}^{--}dU^{\und{m}++}, \qquad A^{IJ}=
U_{\und{m}}^{I}dU^{\und{m}J}.
\end{equation}
Then the worldsheet torsions are defined by
\begin{equation}\plabel{T++}
 T^{++} ={\cal D}e^{++} = de^{++} - e^{++} \wedge \om,
\end{equation}
\begin{equation}\plabel{T--}
 T^{--} ={\cal D}e^{--} = de^{--} + e^{--} \wedge \om,
\end{equation}
\begin{equation}\plabel{T+q}
 T^{+q} ={\cal D}e^{+q} = de^{+q}
 - {1\over 2}  e^{+q} \wedge \om +
 {1\over 4}  e^{+p} \wedge A^{IJ}
 \g^{I}_{p\dot{p}} \g^{J}_{q\dot{p}}
\end{equation}
and {\sl can be calculated directly} from Eqs. \p{4.2.8} with the use of 
Eqs. \p{2.3}, \p{2.7}, \p{2.15}, \p{2.16} and the superembedding
equation
\p{4.11}.

To proceed in this way, one shall use the general decomposition
for the remaining  fermionic supervielbein forms
\begin{equation}\plabel{E-}
\hat{E}^{-}_{\dot{q}}=
e^{+q}a^{--}_{q\dot{q}}+
e^{++}\psi^-_{++\dot{q}}+
e^{--}\psi^-_{--\dot{q}},
\qquad \end{equation}
as well as for the covariant Cartan forms \p{2.14}, e.g.
\begin{equation}\plabel{f++}
f^{++I} =
e^{+q}
f^{++I}_{+q}+
e^{++}
f^{++I}_{++} +
e^{--}
f^{++I}_{--}.
\qquad \end{equation}

As a result of calculations one arrives at the standard constraints for
the bosonic torsion \p{T++}
\begin{equation}\plabel{T++=}
 T^{++} = 2i e^{+q} \wedge e^{+q}.
\end{equation}
The expressions for the torsion forms \p{T--}, \p{T+q} are more
complicated.
On the surface of the superembedding condition \p{4.11} they can be
written as
\begin{equation}\plabel{T--=}
 T^{--} = -2i E^{-}_{\dot{q}} \wedge E^{-}_{\dot{q}},
\end{equation}
\begin{equation}\plabel{T+q=}
 T^{+q} =
 {1\over 2}
 E^{-}_{\dot{q}} \wedge f^{++I} \g^{I}_{q\dot{q}},
\end{equation}
where $E^{-}_{\dot{q}}$ and  $f^{++I}$
are  defined in Eqs. \p{E-}, \p{f++}.
The curvature and gauge field strength are determined from the
Maurer--Cartan equations  \p{G}, \p{R}
\begin{equation}\label{R1}
d \omega = {1\over 2} f^{--I} \wedge f^{++I},
\end{equation}
$$
 {\cal F}^{IJ}  \equiv d A^{IJ} + A^{IK} \wedge A^{KJ}
= - f^{--[I} \wedge f^{++J]}.
$$

The constraints are essentially simplified  when all the superfield
equations of motion are taken into account (see below).

\bigskip

\section{Investigation of superfield equations}

It is convenient to begin the complete investigation of the superfield
equations \p{4.11}, \p{4.22}--\p{4.21} with the study of the
integrability conditions \p{3.5''} of the superembedding equation
\p{4.11}. To this end we assume that the  worldsheet geometry induced by 
embedding in accordance with Eqs. \p{4.2.8}, \p{cCf} and substitute the
most general expression \p{E-} for remaining fermionic supervielbein
forms.
The result can be decomposed into the basic 2--forms
$e^{+q} \wedge e^{+p}$,
$e^{\pm\pm} \wedge e^{+p}$,
$e^{++} \wedge e^{--}$.
The coefficients $({\cal D}E^I)_{+q+p}$, $({\cal D}E^I)_{+q\pm\pm}$,
$({\cal D}E^I)_{--~++}$, which emerge in the product with basic
2--forms,
are superfield equations.
It is convenient to classify 
them by the dimensionality of the basic forms
$[e^{+q} \wedge e^{+p}]=1$,
$[e^{\pm\pm} \wedge e^{+p}]=3/2$,
$[e^{++} \wedge e^{--}]=2$.

At dimension $1$ one gets
\begin{equation}\plabel{DEI1}
({i\over 4} {\cal D}E^I)_{+q+p}= a_{(p \dot{q}} \g^{I}_{q)\dot{q}}=0.
\end{equation}
Substituting the most general decomposition
$a_{q \dot{q}}
=a^I \g^I_{q \dot{q}} + a^{J_1J_2J_3} \g^{J_1J_2J_3}_{q \dot{q}}$ one
can
obtain an equivalent form of Eq. \p{DEI1}
\begin{equation}\plabel{DEI2}
a^I \d_{qp}  + a^{J_1J_2J_3} \g^{IJ_1J_2J_3}_{qp}=0,
\end{equation}
which evidently implies $a^I=0$, $a^{J_1J_2J_3}=0$. Hence
the general solution of Eq. \p{DEI1} is trivial
\begin{equation}\plabel{a=0}
a_{q \dot{q}}=0.
\end{equation}
Thus a consequence of superembedding equation \p{4.11}
is that the fermionic supervielbein 1--form $E^{-}_{\dot{q}}$ has the
form \p{3.6'}, or, equivalently
\begin{equation}\plabel{E-2}
\hat{E}^{-}_{\dot{q}}= e^{++}
\psi^{~~~-}_{++\dot{q}}+
e^{--}
\psi^{~~~-}_{--\dot{q}}.
\qquad
\end{equation}
The equations of dimensions $3/2$ and $2$ provide us with the
expressions
for the covariant Cartan forms \p{2.14}
\begin{equation}\plabel{f++I}
f^{++I} = -4i e^{+q} \g^I_{q\dot{q}}
\psi^{~~~-}_{--\dot{q}}+ e^{++}h^I +
e^{--}
f^{++I}_{--},
\qquad
\end{equation}
\begin{equation}\plabel{f--I}
f^{--I} = -4i e^{+q} \g^I_{q\dot{q}}
\psi^{~~~-}_{++\dot{q}}+
e^{++}
f^{--I}_{++} +
e^{--}h^I.
\qquad
\end{equation}
(Actually the simplest way to derive \p{f++I}, \p{f--I} is to substitute
\p{E-2} back into Eq. \p{3.5''}).

 \bigskip

It is instructive to find the expressions for the left--hand--sides of
bosonic and fermionic superfield equations \p{3.6}, \p{3.5}, which
we obtained in the frame of generalized action principle. They are
\begin{equation}\plabel{MI}
\hat{M}_2^I=
e^{++}\wedge e^{--}h^I +
4i e^{--}\wedge e^{+q}
\psi^{~~~-}_{--\dot{q}}\g^{I}_{q\dot{q}},           \qquad
\end{equation}
\begin{equation}\plabel{Psi}
\hat{\Psi}_2{}^{-}_{\dot{q}} = e^{++} \wedge e^{--}
\psi^{~~~-}_{--\dot{q}}. \qquad
\end{equation}
The leading component $~~h^I_0 = h^I\vert_{\eta =0}$ of the superfield $h^I$
is known as the mean curvature of the 
worldsheet. The bosonic equations is essentially $h^I_0=0$ and the
fermionic ones are
$(\psi^{~~~-}_{--\dot{q}})_0\equiv
\psi^{~~~-}_{--\dot{q}}
\vert_{\eta =0}=0$.
However, by now we have not obtained them from the action.

 \bigskip

Substituting \p{E-2}, \p{f++I}, \p{f--I} into Eqs. \p{T--=}, \p{T+q=}
one
arrives at the torsion constraints for induced worldsheet supergravity
on the surface of superembedding equation \p{4.11}
\begin{equation}\plabel{T++=1}
 T^{++} = 2i e^{+q} \wedge e^{+q},
\end{equation}
\begin{equation}\plabel{T--=1}
 T^{--} = -4i
 e^{++} \wedge e^{--}
\psi^{~~~-}_{++\dot{q}}
\psi^{~~~-}_{--\dot{q}},
\end{equation}
\begin{equation}\plabel{T+q=1}
 T^{+q} =  2i
 e^{\pm\pm} \wedge e^{+q}
 \g^I_{q\dot{q}}
\psi^{~~~-}_{\pm\pm\dot{q}} +
 e^{++} \wedge e^{--}
\left( f^{++I}_{--}
 \g^I_{q\dot{q}}
\psi^{~~~-}_{++\dot{q}} - h^I
 \g^I_{q\dot{q}}
\psi^{~~~-}_{--\dot{q}}
\right).
\end{equation}
After making substitution of Eq. \p{f++I},  
the lowest dimensional component of the Peterson-Codazzi equation
\p{PC} results in
\begin{equation}\plabel{66.1}
 \g^{I}_{(q\dot{q}}
 {\cal D}_{+p)} \psi^{~~~-}_{--\dot{q}} =
 - {1\over 2} \d_{qp} h^I
\end{equation}
Eq. \p{66.1} implies the dependence of the bosonic superfield equation of
motion which follows from the generalized action on the fermionic equation.
It can be 'solved' algebraically by (cf. \p{3.61}) 
\begin{equation}\plabel{66.2}
 {\cal D}_{+p} \psi^{~~~-}_{--\dot{q}} =
 -{1\over 2} \g^{I}_{q\dot{q}} h^I.
\end{equation}

\section{Noether identities and hidden gauge symmetries}

To carry out the analysis of dynamical degrees of freedom one needs
to know all the gauge symmetries. From the standard approach
\cite{GS91,GHDS} it is known that
there should be some hidden gauge symmetries acting on the Lagrange
multiplier
 superfields which reduce their dynamical content.
However, the analysis of gauge symmetries inherent to a superfield
action
is rather involved.

To overcome the cumbersome calculations we propose to use the second
Noether
theorem. It states that any gauge symmetry of the action results
in a dependence of equations of motion, and that the converse
is true as well,
i.e. {\sl an  interdependence of equations of motion indicates the
presence
of a gauge symmetry}.

In our case the second  Noether theorem allows to state that the
superfield
action \p{4.1} possesses the gauge symmetries with basic
terms in the transformation low defined by
\begin{equation}\plabel{5.18}
\d P^{NM}=(-)^K\partial_K\Sigma^{KMN},
\end{equation}
\begin{equation}\plabel{5.19}
\d P_I^{M}=2(-)^N\partial_NS^{+q+pI} E^{~N}_{+q}E^{~M}_{+q}(-)^N+ 
\ldots
\qquad
\end{equation}
where $\ldots$ denotes the terms dependent either on
the superfield parameter
$\Sigma^{KMN}= -(-)^{MN} \Sigma^{KNM}=
\Sigma^{[KMN)}$ (graded--antisymmetric supertensor)
or on the parameter
$S^{+q+pI}$ which obey the properties
\footnote{The terms dependent on $S^{+q+pI}$ in \p{5.19} involve it in
products with different superfields and are unessential
for the analysis of dynamical degrees of freedom as the latter can
be carried out in linear approximation.}
\begin{equation}\plabel{5.21}
{\tilde S}^{+q+pI}={\tilde S}^{+p+qI},   \qquad
{\tilde S}^{+p+qI}\g^I_{q\dot{q}} =0,   \quad \Rightarrow \quad
{\tilde S}^{+q+qI}=0.
\end{equation}

The dependence \p{3.7} of the integrability conditions
\p{4.13} for the equation \p{4.12}
\begin{equation}\plabel{8.1}
E^I=0, \quad ({\cal D}E^I)_{+q+p}=0, \qquad  \Rightarrow \qquad
d{\cal L}_2=0,
\end{equation}
mentioned in Section 3, indicates the presence of the symmetry
with basic transformation low \p{5.18}.
Indeed, if one considers the variation of the action \p{4.1} under the
Lagrange multiplier transformations \p{5.18}, one easily finds
(cf. \p{4.10})
\begin{equation}\plabel{5.2}
\d_0 S=-1/3 \int d^2\xi
d^2\eta \Sigma^{MNK} (d{\cal L}_2)_{KNM}
\end{equation}
But $d{\cal L}_2$ vanishes as a result of
Eq. \p{4.11}, which emerges as a result of the variation of the
Lagrange multiplier $P_I^M$, and of its integrability condition
 \p{8.1} only. This guarantees that {\bf i)} one can find
transformations
for the Lagrange multiplier superfield  $P_I^M$ which will compensate
the variation \p{5.2}, and that {\bf ii)} other superfields can be
regarded to
be inert under the  gauge symmetry \p{5.18}.

\bigskip

To establish the presence of the gauge symmetry \p{5.19} one can turn
back to
the lowest dimensional component \p{DEI1} of the integrability condition
\p{3.5''} for Eq. \p{4.11}. Eq. \p{DEI1} carries a
{\bf 36}$_s \times$ {\bf 8}$_v=$ {\bf 288} reducible representation
of the $SO(8)$ group. Its decomposition onto irreducible representations
reads
\begin{equation}\plabel{288}
\hbox{\bf 288}= \hbox{\bf 8} + \hbox{\bf 56} + \hbox{\bf 224},
\qquad \Leftrightarrow \qquad
  {}^{\fbox{}}_{\fbox{}}\otimes {}^{\fbox{}} ~=~
 {}^{\fbox{}}~+ ~{}^{\fbox{}\fbox{}\fbox{}} ~+~
{}^{\fbox{}\fbox{}\fbox{}\fbox{}}_{\fbox{}}
\end{equation}
or more explicitly
\begin{equation}\plabel{288E}
({\cal D}E^I)_{+q+p}= s^I \d_{qp} + s^{I_1I_2I_3} \g^{II_1I_2I_3}_{qp}+
s^{I,I_1I_2I_3I_4} \g^{I_1I_2I_3I_4}_{qp}
\end{equation}
where
\begin{equation}\plabel{224}
s^{I,J_1J_2J_3J_4} = s^{J_1,IJ_2J_3J_4}=
{1\over 4!}\E^{J_1J_2J_3J_4I_1I_2I_3I_4} s^{I,I_1I_2I_3I_4}
\equiv  {}^{{\fbox{}\fbox{}\fbox{}\fbox{}}}_{{\fbox{}}}  \qquad
\end{equation}
is the $SO(8)$ tensor description of the {\bf 224} irreducible
representation.

One can see that to arrive at the trivial solution \p{a=0} it is enough
to
consider only {\bf 8} and {\bf 56} irreducible parts of Eq.\p{DEI1}:
\begin{equation}\plabel{a=0i}
s^{I}=\propto ({\cal D}E^I)_{+q+q}=0, \quad
s^{J_1J_2J_3}=\propto  ({\cal D}E^I)_{+q+p} \g^{IJ_1J_2J_3}_{qp} =0
\qquad \Rightarrow \qquad
a_{q \dot{q}}=0.
\end{equation}
Thus the {\bf 224} irreducible part of Eq. \p{DEI1} is satisfied
identically
due to  {\bf 8} and {\bf 56} irreducible parts of this equation.
This is the Noether identity for the gauge symmetry \p{5.19} with
\begin{equation}\plabel{SSG}
S^{+q+pI}=
S^{I,I_1I_2I_3I_4} \g^{I_1I_2I_3I_4}_{qp}, \qquad
\end{equation}
\begin{equation}\plabel{S224}
 S^{I,J_1J_2J_3J_4} = S^{J_1,IJ_2J_3J_4}=
{1\over 4!}\E^{J_1J_2J_3J_4I_1I_2I_3I_4} S^{I,I_1I_2I_3I_4}
\equiv \hbox{{\bf 224}}
\equiv  {}^{\fbox{}\fbox{}\fbox{}\fbox{}}_{\fbox{}} , \qquad
\end{equation}
Eq. \p{SSG} provides the general solution of the
gamma--tracelessness conditions for the vector--spin-tensor $S^{+q+pI}$
\p{5.21}.

Hence, using the second Noether theorem we have proved that the
superfield
action \p{4.1} possesses hidden gauge symmetries \p{5.18}, \p{5.19}.

\bigskip

\section{The fate of Lagrange multiplier superfields and component
equations of motion}

The gauge symmetry \p{5.18} and the
equation of motion \p{4.22} imply that the Lagrange
multiplier
superfield $P^{MN}$ do not carry dynamical degrees of freedom.
Indeed, the  general solution of Eq.\p{4.22} has the form \cite{GHDS}
($(\eta )^8 \equiv  {1/8!} \e_{q_1\ldots q_8} \eta^{q_1} \ldots \eta^{q_8}
$ \p{n}) 
\begin{equation}\plabel{6.1}
P^{NM}=(-)^K\partial_K{\tilde \Sigma}^{KNM}+ \d^N_n\d^M_m\e^{mn}
(\eta)^8 ~T,   \qquad
dT=0. \qquad
\end{equation}
The  first term in Eq.\p{6.1} can be gauged away by the transformations
Eqs.\p{5.18}, \p{5.19} with parameter
$\Sigma^{KNM}=-{\tilde \Sigma}^{KNM}$. In this gauge one gets
\begin{equation}\plabel{6.2}
P^{NM}= \d^N_n\d^M_m\e^{mn}   (\eta )^8 ~T,
\qquad dT=0 \qquad
\end{equation}
and finds that the constant
$T$ is the string tension.

The situation with the Lagrange multiplier $P^M_I$
is slightly more complicated.
The general solution of Eqs.\p{4.15}, \p{4.16} is
\begin{equation}\plabel{6.3}
P^{M}_I=P^{+q}_IE_{+q}^{~M}.  \qquad
\end{equation}

\bigskip

At this point it is pertinent to note that 
Eq. \p{4.21} is satisfied identically when Eqs. \p{6.3} and \p{E-2} are
taken into account. The identification \p{4.2.8} makes evident that this
dependence reflects  $n=(8,0)$ {\sl local
worldsheet supersymmetry} on the language of Noether identities.
The bosonic reparamterization is reflected by the fact that
Eqs. \p{4.18}, \p{4.19} are dependent on Eqs. \p{E-2}, \p{6.3}
and \p{4.20}.

\bigskip

In the gauge \p{6.2} the Eq. \p{4.20}  acquires the form
\begin{equation}\plabel{4.201}
P^{+q}_I\g^I_{q\dot{q}}= -4 T (\eta )^8 e(\zeta)
\psi^{~~~-}_{--\dot{q}}
\end{equation}
where
\begin{equation}\plabel{e(zeta)}
e(\zeta) = e(\xi, \eta )= {1 \over 2} \e^{mn} e^{++}_m e^{--}_n,
\end{equation}
and Eq. \p{4.17} becomes
\begin{equation}\plabel{4.171}
(-)^M(\partial_M (P^{+q}_I E_{+q}^{M}) + A^{IJ}_{+q} P^{+q}_J)=
2 (\eta)^8 ~T ~e ~\tilde{h}^I,
\end{equation}
where
\begin{equation}\plabel{thI}
\tilde{h}^I =h^I + {2i \over e} e^{mn} e^{--}_m e^{+q}_n
\psi^{~~~-}_{--\dot{q}}
\end{equation}

Eqs. \p{4.171}, \p{4.201} and the gauge symmetry \p{5.19} with the
parameter \p{5.21} imply that
\\ {\bf i)}
the Lagrange multiplier superfield $P^M_I$ does not contain
independent dynamical degrees of freedom,
\\ {\bf ii)} the component coordinate equations of motion for
superstring
\begin{equation}\plabel{eqmf}
(\psi^{~~~-}_{--\dot{q}})_0 =0,
\end{equation}
\begin{equation}\plabel{eqmb}
(h^I)_0 =0
\end{equation}
are satisfied.

We prove this fact in the Appendix.
Similar mechanism was discovered for the  first time in Ref.
\cite{PT} where  superfield actions for supermembrane in
D=4,5,7 was constructed and an STV-like action for a 2--dimensional
extended object in D=11 was considered.

Thus we conclude that the superfield action Eq.\p{4.1} indeed describes
the $D=10,~~N=1$ closed Green-Schwarz superstring.

\bigskip

\section*{Conclusion}

In this paper we constructed a new version of superfield action for
the $D=10$, $N=1$ closed superstring (i.e. a heterotic superstring
without heterotic fermions).
The action possesses manifest $n=(8,0)$ worldsheet supersymmetry and
involves
Lorentz harmonics as auxiliary superfields.
The second ('Wess-Zumino' \cite{GHDS}) term of the action is constructed
from the Lagrangian form of the generalized action principle \cite{bsv}.

To find the hidden gauge symmetries which allow one to
remove redundant degrees of freedom from the Lagrange Multiplier
superfields we used the second Noether theorem.

In distinction to the original action \cite{GHDS} and to the action from
\cite{bcsv} our functional does not contain worldsheet supervielbein
$e_M^{~A}$ explicitly. Thus our action can be considered as a
superfield counterpart
of the Nambu--Goto action while the original one \cite{GHDS} should be
regarded as a counterpart of the
Brink--Di Vecchia--Howe--Polyakov
functional.
This property is important because the worldsheet supervielbein
should be either constructed from 'prepotentials' or
subject to a set of torsion constraints. This is not a problem for
the $N=1$ superstring, but could produce some difficulties for the
construction of superfield actions for other branes with high
dimensional
worldsheet $p>1$ and $n\leq 8$ worldsheet supersymmetries.
(E.g. the off--shell description of $d=4$, $n=2$ superfield supergravity
 is impossible without the use of harmonic variables \cite{GIKOS}).

 To deal with the new version of superfield action \p{4.1} we do not
need any assumptions about
 worldsheet supergravity. Instead we have to assume that some components
 \p{4.7.2} of the pull--back of the supervielbein of the target
superspace
 form an invertible $10 \times 10$ supermatrix
 (cf. \p{4.7.4})
\begin{equation}\plabel{7.1}
\hat{E}^{~A}_M=
(\hat{E}^{++}_M, \hat{E}^{--}_M, \hat{E}^{+q}_M), \qquad sdet(
\hat{E}^{~A}_M)\not= 0.
\end{equation}
In such a way we actually perform (implicitly) an identification
of the form $\hat{E}^{A} = d\zeta^M E_M^{~A}\left(\hat{Z}(\zeta)\right)$ 
with a worldsheet supervielbein $e^{A} = (e^{++}, e^{--}, e^{+q})$.
The worldsheet spin connection $\om_M$ and $SO(8)$ gauge fields
$A_M^{IJ}$ can be
constructed from Lorentz harmonic superfields (and target superspace
spin
connections when the general $D=10$, $N=1$ supergravity background is
considered).

Thus the geometry of the worldsheet superspace
is induced by embedding in accordance with the following rules
\begin{equation}\plabel{7.2}
e^{\pm\pm} =
\hat{E}^{{\pm}{\pm}}\equiv
\hat{\Pi}^{\und{m}}U_{\und{m}}^{{\pm}{\pm}},
\qquad
e^{+q} =
\hat{E}^{+q}_M \equiv
d\hat{\Th}^{\und{\mu}}V_{\und{\mu}q}^{~+},
\end{equation}
$$
\om = {1 \over 2} U_{\und{m}}^{--}dU^{\und{m}++}, \qquad
A^{IJ}= U_{\und{m}}^{I}dU^{\und{m}J}.
$$

The worldsheet supergravity constraints can be obtained as integrability 
conditions for Eqs. \p{7.2} taken on the surface of superembedding
equations and have the form
\p{T++=1},
\p{T--=1},
\p{T+q=1}.
Such a mechanism of the generation of worldsheet supergravity
constraints
is characteristic of the generalized action \cite{bsv}.
Its bosonic counterpart was used in \cite{baku} in an early
consideration
of the Brane--World scenario.

Note that the new formulation provides as well a 'complete twistorization' of
$N=1$ (heterotic) superstring in a way alternative to the one proposed in Ref.
\cite{bcsv}. Indeed, the basic superembedding equation \p{4.11}
and Eq. \p{7.2} imply that 
$$
 {\Pi}^{\und{m}}={1\over 2}  e^{++} U^{\und{m}--}+ 
{1\over 2}  e^{--} U^{\und{m}++}
$$ 
Thus we arrive at
$$
{\Pi}^{\und{m}}_{++} ={1\over 2} U^{\und{m}--}\equiv {1\over 16}
V^{-}_{\dot{q}}\tilde{\G}_{\und{m}}
V^{-}_{\dot{p}}, \qquad 
{\Pi}^{\und{m}}_{--} ={1\over 2} U^{\und{m}++}\equiv {1\over 16}
V^{+}_{{q}}\tilde{\G}_{\und{m}}
V^{+}_{{p}}.
$$
These relations provide a twistor--like solution for the 
Virasoro constraints (cf. \p{2.4'}, \p{2.17}) 
$ {\Pi}^{\und{m}}_{++} {\Pi}_{++\und{m}} =0$,  
${\Pi}^{\und{m}}_{++} {\Pi}_{++\und{m}} =0$.  

\bigskip 

The methods developed in this paper should simplify the investigation
of superfield actions for higher superbranes in low dimensions
and can be useful for studying
 the physical contents of 'spinning superbrane' models in
$D=11$ and $D=10$ type II superspaces
which are described  by STV-like actions with $16$ worldvolume
supersymmetries  \cite{PT,BS}.

\section*{Acknowledgments} 

The authors are grateful to D. Sorokin for useful considerations and
helpful remarks on reading the manuscript 
and J. Lukierski for useful communications. 
One of us (I.B.) would like to thank Mario Tonin for valuable discussions 
and the Padova section of INFN for the hospitality at the Padova University 
during the early stages of  this work. T.B. would like to thank the 
Abdus Salam ICTP for the hospitality extended to her during her visit when 
the main part of this work  was done.

\newpage

\section*{Appendix }

Here we present some details of a direct derivation of the gauge
symmetry
\p{5.19} and use it
to prove that
\\ i) the Lagrange multiplier superfield $P^M_I$ does not contain
independent dynamical degrees of freedom and that
\\  ii) the component coordinate equations of motion for superstring
\p{eqmf}, \p{eqmb} are satisfied.

 Let us consider the transformations

\begin{equation}\plabel{5.190}
\d P_I^{M}= 2(-)^N{\cal{D}}_NS^{NMI}. \qquad
\end{equation}
where
$\Sigma^{KMN}=\Sigma^{[KMN\}}(\zeta)$ is a
graded--antisymmetric supertensor and
\begin{equation}\plabel{5.1}
{\cal{D}}S^{NMI}\equiv d
S^{NMI}+S^{NMI}A^{JI}=d\zeta^K{\cal{D}}_KS^{NMI}.
\end{equation}
The variation of the action \p{4.1} is
\begin{equation}\plabel{5.4}
\d_2 S= \int d^2\xi {\hat d}^8\eta 2(-)^N{\cal D}_N S^{NMI}E^I_M=
\end{equation}
$$
\int d^2\xi {\hat d}^8\eta (4iS^{MNI}\g^I_{q\dot{q}}
E^{~+}_{N{q}} E^{~-}_{M\dot{q}}(-)^M -
S^{MNI}E^{++}_N f^{--I}_M-S^{MNI}E^{--}_Nf^{++I}_M=
$$
$$
\int d^2\xi {\hat d}^8\eta (4iS^{+q+pI}\g^I_{q\dot{q}}
E^{~M}_{+{q}}(-)^M E^{~-}_{M\dot{q}}(-)^M + \ldots
$$
where $\ldots$ denote the terms independent of $S^{+q+pI}=
S^{MNI} E^{~+}_{N{q}} E^{~+}_{Mp}(-)^M$.
Thus the gamma-traceless part of $S^{+q+pI}$ indeed disappears from
the action variation and thus can be associated with a parameter
of a gauge symmetry.

To find dynamical equations and to analyze the dynamical content
of the Lagrange multiplier $P_I^{+q}$ let us turn to
 Eqs. \p{4.171}, \p{4.201}.
For the sake of simplicity, we turn to the linear approximation
where Eqs. \p{4.171}, \p{4.201}  acquire the form
\begin{equation}\plabel{4.172}
D_{+q} P^{+q}_I =
2 (\eta)^8 T ~h^I, 
\end{equation}
\begin{equation}\plabel{4.202}
P^{+q}_I\g^I_{q\dot{q}}= -4 T (\eta )^8
\psi^{~~~-}_{--\dot{q}}
\end{equation}
and the gauge symmetry \p{5.19} reduces to 
\begin{equation}\plabel{5.201}
\d P^{+q}_I =  D_{+p} {S}^{+p+q I}, \qquad
\end{equation}
In Eqs. \p{4.172}, \p{5.201} $D_{+q}$
is the flat superspace covariant derivative \p{D+q} and 
$$
{S}^{+p+q I}- {S}^{+q+p I}= {S}^{+p+q I}
\g^{I}_{q\dot{q}}
= \tilde{S}^{+p+p I}=0.
$$
carries {\bf 224} irreducible representation of $SO(8)$ \p{S224}.

The linearized version of superfield bosonic and fermionic equations
\p{eqmb}, \p{eqmf} is
$$
h^I = \propto
\partial_{++}
\partial_{--} X^I, \qquad
\psi^{~~~-}_{--\dot{q}} =
\propto \partial_{--}
\Theta^{-}_{\dot{q}}.
$$
Eqs. \p{66.2} can be used to write the decomposition of the superfield
$\psi^{~~~-}_{--\dot{q}}$
\begin{equation}\plabel{psi2}
\psi^{~~~-}_{--\dot{q}} =
(\psi^{~~~-}_{--\dot{q}})_0 (\xi ) - {1 \over 2}
\g^I_{q\dot{q}} h^I_0 + \ldots
\end{equation}

To proceed further, the following identities are useful
(see Eqs. \p{n} for definitions)
\begin{equation}\plabel{D8}
 D_{+q} (\eta )^8 = (\eta )^7_{q}, \quad
 D_{+p} (\eta )^7_{q}
 = (\eta )^6_{qp}+ 2i (\eta )^8  \d_{qp} \partial_{++}, \quad
 D_{+q} (\eta )^7_{q}=
 16i (\eta )^8 \partial_{++}, 
\end{equation}

The general solution of Eq. \p{4.202} has the form
\begin{equation}\plabel{4.202s}
P^{+q}_I = \tilde{P}^{+q}_I  - {1 \over 2}
(\eta)^8 ~T ~\psi^{~~~-}_{--\dot{q}}, \qquad
\end{equation}
where
$\tilde{P}^{+q}_I$ is gamma--traceless
\begin{equation}\plabel{4.203s}
\tilde{P}^{+q}_I  \g^I_{q\dot{q}}=0.
\end{equation}
Substituting \p{4.202s} into Eq. \p{4.171} one arrives at the
following equation for
$\tilde{P}^{+q}_I$
$$
D_{+q} \tilde{P}^{+q}_I = {1 \over 2}
(\eta )^7_q
\g^{I}_{q\dot{q}}
(\psi^{~~~-}_{--\dot{q}})_0
+
2 (\eta)^8 T ~h^I_0
$$
However, using the decomposition \p{psi2} one can collect two terms
of the above equation into the first one, but with the component
$(\psi^{~~~-}_{--\dot{q}})_0$ replaced by the superfield
$\psi^{~~~-}_{--\dot{q}}$.
\begin{equation}\plabel{4.173}
D_{+q} \tilde{P}^{+q}_I = {1 \over 2}
(\eta )^7_q
\g^{I}_{q\dot{q}} \psi^{~~~-}_{--\dot{q}}
\end{equation}
The general solution of Eqs. \p{4.173} for {\sl gamma-traceless}
$\tilde{P}^{+q}_I $ \p{4.203s} is a sum of the general solution of the
homogeneous equation and a particular solution of
the inhomogeneous one
$$
  \tilde{P}^{+q}_I =
  (\tilde{P}^{+q}_I )_{gen} +
  (\tilde{P}^{+q}_I )_{par},
  \qquad
  D_{+q} (\tilde{P}^{+q}_I )_{gen} = 0.
$$
The former has the form
$$
  (\tilde{P}^{+q}_I )_{gen} = D_{+p} {\S}^{+p+q I}, \qquad
$$
with the parameter ${\S}^{+p+q I}$ satisfying
$$
\S^{+p+q I}- \S^{+q+p I}= \S^{+p+q I}
\g^{I}_{q\dot{q}}
= \S^{+p+p I}=0.
$$
It can be gauged away by the symmetry \p{5.201}.
Thus the solution of the equation \p{4.173} does not contain 
any indefinite constants. Hence, just at this point, one can conclude
that the Lagrange multiplier $P_I^M$ does not contain any dynamical
degree of freedom.

The general solution of \p{4.173}  thus reduces to a particular
solution. The latter can be easily obtained using the
decomposition  of the superfield in the components
\begin{equation}\plabel{Ppart}
\tilde{P}^{+q}_I =
-{i \over 2} (\eta )^7_q T {1 \over \partial_{++}}h^I_0 +
{1 \over 2} (\eta )^8 T \g^{I}_{q\dot{q}} (\psi^{~~~-}_{--\dot{q}})_0
\end{equation}
However, the solution is not traceless.
When we substitute \p{Ppart} in the tracelessness conditions
\p{4.203s}, we find that
$$
h^I_0=0, \qquad       (\psi^{~~~-}_{--\dot{q}})_0  =0,
\qquad
\tilde{P}^{+q}_I = 0.
$$
In other words, there is no a solution  of Eq. \p{4.173} with traceless
${P}^{+q}_I$.

Thus the Lagrange multiplier
$P^M_I$ can be gauged away and the component bosonic and fermionic
coordinate equations \p{eqmb}, \p{eqmf} emerge as a result of Eqs.
\p{4.171}, \p{4.201}.

Note that local $n=(8,0)$ worldsheet supersymmetry requires that if
the leading component of a superfield is equal to zero, then the whole
superfield is equal to zero as well (cf. \cite{ST}).
Thus the {\sl supersymmetric} solution
of Eqs. \p{4.171}, \p{4.201} corresponds to superfield  equations
\p{3.5}, \p{3.6}, which are produced by the
generalization action.

\end{document}